\documentclass[10pt,english,aps,prd,10pt,notitlepage,preprintnumbers]{revtex4-1}
\usepackage[T1]{fontenc}
\usepackage[latin9]{inputenc}
\usepackage{bm}
\usepackage{amstext}
\usepackage{amssymb}
\usepackage{esint}

\makeatletter

\@ifundefined{textcolor}{}
{%
 \definecolor{BLACK}{gray}{0}
 \definecolor{WHITE}{gray}{1}
 \definecolor{RED}{rgb}{1,0,0}
 \definecolor{GREEN}{rgb}{0,1,0}
 \definecolor{BLUE}{rgb}{0,0,1}
 \definecolor{CYAN}{cmyk}{1,0,0,0}
 \definecolor{MAGENTA}{cmyk}{0,1,0,0}
 \definecolor{YELLOW}{cmyk}{0,0,1,0}
}

\def\Mpl{M_{\mathrm{Pl}}}

\usepackage{babel}

\makeatother

\usepackage{babel}
\begin{document}

\preprint{YITP-14-99}

\title{Ghosts in classes of non-local gravity}

\author{Antonio De Felice}

\affiliation{Yukawa Institute for Theoretical Physics, Kyoto University, 606-8502,
Kyoto, Japan}

\author{Misao Sasaki}

\affiliation{Yukawa Institute for Theoretical Physics, Kyoto University, 606-8502,
Kyoto, Japan}

\date{\today}
\begin{abstract}
We consider a class of non-local gravity theories where the Lagrangian
is a function of powers of the inverse d'Alembertian operator acting
on the Ricci scalar. We take an approach in which the non-local Lagrangian
is made local by introducing auxiliary scalar fields, and study the
degrees of freedom of the localized Lagrangian. We find that among
the auxiliary scalar fields introduced, some of them are always ghost-like.
That is, in the Einstein frame they develop a negative kinetic term.
Because of this, except for a particular case already known in the
literature, in general, it is not clear how to quantize these models
and how to interpret this theory in the light of standard field theory. 
\end{abstract}
\maketitle

\section{Introduction}

Among the theories introduced to describe the late-time acceleration
of the universe, the modified-gravity paradigm has attracted much
interest, because it explicitly states that the reason for the acceleration
of the universe is due to a modified gravity law which is mostly felt
at very large scales. The exploration of different ways of modifying
gravity have started since the pioneeristic works in the so-called
$f(R)$ gravity. Many other theories have been proposed since then.
Among others, let us mention a few of them here: the extension of
$f(R)$ theories to $f(R,G)$ theories where $G$ stands for the Gauss-Bonnet
term, the DGP model motivated by the possible existence of spatial
extra-dimensions, Galileon theories and general scalar-tensor theories
of the Horndeski Lagrangian with second order differential equations.
All these theories generalize the Einstein-Hilbert Lagrangian by introducing
second order Lagrangians (or to Lagrangians which reduce to them,
as in the $f(R)$ case) for gravity and some extra scalar degrees
of freedom. More recently a new class of modifications of gravity
has been introduced, so-called non-local theories of gravity. The
Lagrangian of these theories consists of terms which are non-local
in the form $f(\cdots,\Box^{-1}R,\cdots)$ \cite{Deser}. These theories
have attracted some attention both theoretically \cite{Dirian,Dvali,Foffa,Calcagni,Capozziello,Cognola,Deffayet,Joukov,Kehagias,Kluson,Koivisto,Koivisto-2,Koshelev,Maggiore,Mancarella,Sasaki,Simon,Woodard,Biswas}
and phenomenologically \cite{Barreira,Bamba,Jhingan,Liu,Milani,Nogiri},
as a possible alternative to dark energy that renders the universe
accelerated at late times.

How to deal with this kind of Lagrangian is a non-trivial topic. We
will consider here the case of a general function studied recently
in the literature \cite{Dirian-1} 
\begin{equation}
\mathcal{L}=\sqrt{-g}\, f(R,\Box^{-1}R,\cdots,\Box^{-n}R)\,,\qquad\mathrm{with}\qquad n<+\infty\,,\label{eq:LagOne}
\end{equation}
and we will try to understand its content. The meaning of such terms
in the Lagrangian is obscure, and not very well understood. Some people
take the point of view (see e.g.\ \cite{Foffa}), that, in order
to make it sensible, the $\Box^{-1}$ operator, must be replaced by
the operator $\Box_{\mathrm{ret}}^{-1}$ where the subscript ``$\mathrm{ret}$''
means the retarded boundary condition. This point of view is non-standard
in the conventional context of variational calculus where setting
initial data is a defining constituent of a theory at level of the
action. Further it is unconformable to the usual quantization procedure
known for known theories based on the Lagrangian formalism.

In this paper, we take a different approach. Pursuing the standard
picture of classical/quantum field theory, we interpret the non-local
Lagrangian (\ref{eq:LagOne}) as equivalent to another, local Lagrangian
which can be derived by introducing auxiliary fields. The resulting
Lagrangian can be studied with the usual tools of field theory. Namely
we consider the Lagrangian, 
\begin{equation}
\mathcal{L}=\sqrt{-g}\left[f(\sigma,U_{1},U_{2},\cdots,U_{n})+\frac{\partial f}{\partial\sigma}\,(R-\sigma)+\lambda_{1}(R-\Box U_{1})+\lambda_{2}(U_{1}-\Box U_{2})+\cdots+\lambda_{n}(U_{n-1}-\Box U_{n})\right].\label{eq:LagDue}
\end{equation}

Having the new local Lagrangian (\ref{eq:LagDue}), we can perform
the usual study of the degrees of freedom in the theory. We then find
that such a Lagrangian contains in general $n$ ghost-like propagating
degrees of freedom in any background. Special cases are also studied,
such as the case $\partial^{2}f/\partial\sigma^{2}=0$, separately.
In all these subcases we find a finite number of ghost degrees of
freedom except for the $n=1$ case. These ghosts are unavoidable, in the sense that they cannot be gauged away. Therefore their presence would make these models, in general, unviable, unless one tunes the mass of these modes to values larger than the cut-off of the theory.

This paper is organized as follows. In Sec.~\ref{sec:general} we
rewrite the general non-local Lagrangian in the form of a localized
Lagrangian as given by Eq.~(\ref{eq:LagDue}) and analyze its physical
degrees of freedom. In Sec.~\ref{sec:linear}, we focus on a special
case where the Lagrangian is linear in the Ricci scalar, that is,
the case $\partial^{2}f/\partial\sigma^{2}=0$ in Eq.~(\ref{eq:LagDue}).
Section~\ref{sec:conclusion} is devoted to conclusion and discussions.

\section{General non-local gravity action}

\label{sec:general}

Let us consider the general action, 
\begin{eqnarray}
 &  & S=\int d^{4}x\sqrt{-g}\, f\,;\nonumber \\
 &  & \quad f\equiv f_{1}(R,\Box^{-1}R,\Box^{-2}R,\cdots,\Box^{-n}R)+f_{2}(\Box^{-1}R,\Box^{-2}R,\cdots,\Box^{-m}R)\,,\label{eq:lagrOR-2}
\end{eqnarray}
where $f$ is a general function of $\Box^{-k}R$ ($k=0,1,2,\cdots,\max(n,m)$),
where $n$ and $m$ are positive integers, i.e.\ $1\leq(m,n)<\infty$,
and the function $f_{1}$ is chosen by the condition that it satisfies
\begin{equation}
\frac{\partial^{2}f}{\partial R\partial(\Box^{-n}R)}=\frac{\partial^{2}f_{1}}{\partial R\partial(\Box^{-n}R)}\neq0\,.\label{eq:f1cond}
\end{equation}
Thus $n$ is the largest integer for which this inequality holds.
Note that the choice of $f_{1}$ is not unique, given the function
$f$, but this ambiguity does not affect our discussion below.

As already mentioned in the Introduction,
on allowing ourselves to interpret the action~(\ref{eq:LagOne}) as a
model which can be redefined in terms of a local action
(without e.g.\ assuming the d'Alambertian operators restricted on
particular or prior-given boundary conditions, which would result in
considering different theories),
we can rewrite the action as 
\begin{eqnarray}
S_{m\leq n} & = & \int d^{4}x\sqrt{-g}\Biggl[f_{1}(\sigma,U_{1},U_{2},\cdots,U_{n})+\frac{\partial f_{1}}{\partial\sigma}\,(R-\sigma)+f_{2}(U_{1},\cdots,U_{m})\nonumber \\
 &  & \qquad+\lambda_{1}(R-\Box U_{1})+\lambda_{2}(U_{1}-\Box U_{2})+\cdots+\lambda_{n}(U_{n-1}-\Box U_{n})\Biggr],\label{eq:lagrNEW-2}
\end{eqnarray}
or 
\begin{eqnarray}
S_{m>n} & = & \int d^{4}x\sqrt{-g}\Biggl[f_{1}(\sigma,U_{1},U_{2},\cdots,U_{n})+\frac{\partial f_{1}}{\partial\sigma}\,(R-\sigma)+f_{2}(U_{1},\cdots,U_{m})\nonumber \\
 &  & \qquad+\lambda_{1}(R-\Box U_{1})+\lambda_{2}(U_{1}-\Box U_{2})+\cdots+\lambda_{n}(U_{n-1}-\Box U_{n})+\cdots+\lambda_{m}(U_{m-1}-\Box U_{m})\Biggr].\label{eq:lagrNEW-2b}
\end{eqnarray}
On taking the equations of motion for the fields $\sigma$, and $\lambda_{i}$
($i=1,\cdots,n$), we find 
\begin{eqnarray}
\frac{\partial^{2}f_{1}}{\partial\sigma^{2}}\,(R-\sigma) & = & 0\,,\\
R & = & \Box U_{1}\,,\\
U_{1} & = & \Box U_{2}\,,\\
 & \cdots\nonumber \\
U_{n-1} & = & \Box U_{n}\,,
\end{eqnarray}
for $m\leq n$, and the additional equations, 
\begin{eqnarray}
U_{n} & = & \Box U_{n+1}\,,\\
 & \cdots\nonumber \\
U_{m-1} & = & \Box U_{m}\,,
\end{eqnarray}
for $m>n$. Therefore provided that $\partial^{2}f_{1}/\partial\sigma^{2}\neq0$,
we obtain 
\begin{eqnarray}
\sigma & = & R\,,\\
U_{1} & = & \Box^{-1}R\,,\\
U_{2} & = & \Box^{-1}U_{1}=\Box^{-2}R\,,\\
 & \cdots\nonumber \\
U_{n} & = & \Box^{-1}U_{n-1}=\Box^{-n}R\,,
\end{eqnarray}
for $m\leq n$, and additionally 
\begin{eqnarray}
U_{n+1} & = & \Box^{-1}U_{n}=\Box^{-n-1}R\,,\\
 & \cdots\nonumber \\
U_{m} & = & \Box^{-1}U_{m-1}=\Box^{-m}R\,,
\end{eqnarray}
for $m>n$. We regard the original non-local Lagrangian (\ref{eq:lagrOR-2})
as equivalent to the new one, (\ref{eq:lagrNEW-2}) or (\ref{eq:lagrNEW-2b}).

The importance of the new action, (\ref{eq:lagrNEW-2}) or (\ref{eq:lagrNEW-2b}),
is that it is now clear how many degrees of freedom are present, and
their scalar nature. In fact, we can rewrite them as 
\begin{eqnarray}
S_{m\leq n} & = & \int d^{4}x\sqrt{-g}\Biggl[\left(\frac{\partial f_{1}}{\partial\sigma}+\lambda_{1}\right)R+g^{\alpha\beta}(\partial_{\alpha}\lambda_{1}\partial_{\beta}U_{1}+\partial_{\alpha}\lambda_{2}\partial_{\beta}U_{2}+\cdots+\partial_{\alpha}\lambda_{n}\partial_{\beta}U_{n})\nonumber \\
 &  & {}+f_{1}(\sigma,U_{1},U_{2},\cdots,U_{n})-\sigma\,\frac{\partial f_{1}}{\partial\sigma}+\lambda_{2}U_{1}+\cdots+\lambda_{n}U_{n-1}+f_{2}(U_{1},\cdots,U_{m})\Biggr],
\end{eqnarray}
and 
\begin{eqnarray}
S_{m>n} & = & \int d^{4}x\sqrt{-g}\Biggl[\left(\frac{\partial f_{1}}{\partial\sigma}+\lambda_{1}\right)R+g^{\alpha\beta}(\partial_{\alpha}\lambda_{1}\partial_{\beta}U_{1}+\partial_{\alpha}\lambda_{2}\partial_{\beta}U_{2}+\cdots+\partial_{\alpha}\lambda_{m}\partial_{\beta}U_{m})\nonumber \\
 &  & {}+f_{1}(\sigma,U_{1},U_{2},\cdots,U_{n})-\sigma\,\frac{\partial f_{1}}{\partial\sigma}+\lambda_{2}U_{1}+\cdots+\lambda_{m}U_{m-1}+f_{2}(U_{1},\cdots,U_{m})\Biggr].
\end{eqnarray}

Let us make a field redefinition as 
\begin{equation}
\frac{\partial f_{1}}{\partial\sigma}+\lambda_{1}=\Phi\,,\label{eq:constrGen-1}
\end{equation}
which can be solved for $U_{n}$ provided 
\begin{equation}
\frac{\partial^{2}f_{1}}{\partial\sigma\partial U_{n}}\neq0\,.\label{eq:inequo}
\end{equation}
which is guaranteed by definition, as given by Eq.~(\ref{eq:f1cond}).
Notice that Eq.\ (\ref{eq:inequo}), or, in our approach, its equivalent form (\ref{eq:f1cond}), excludes General Relativity in this class of theories. Therefore the set of theories considered here, are those ones for which it is possible to solve Eq.\ (\ref{eq:constrGen-1}) in terms of the field $U_n$. 
In fact, the field $U_{n}$ becomes a function of the other $n+2$ fields
as 
\begin{equation}
U_{n}=U_{n}(\sigma,U_{j},\Phi-\lambda_{1})\,;\quad j=1,\cdots,n-1\,.
\end{equation}
In this section, we also assume $\partial^{2}f_{1}/\partial\sigma^{2}\neq0$.
The particular case $\partial^{2}f_{1}/\partial\sigma^{2}=0$ will
be discussed separately in the next section.

We find, on differentiating the constraint (\ref{eq:constrGen-1}),
that 
\begin{equation}
\frac{\partial^{2}f_{1}}{\partial\sigma^{2}}\, d\sigma+\frac{\partial^{2}f_{1}}{\partial\sigma\partial U_{j}}\, dU_{j}+\frac{\partial^{2}f_{1}}{\partial\sigma\partial U_{n}}\, dU_{n}+d\lambda_{1}-d\Phi=0\,.
\end{equation}
Recalling that $U_{n}$ is a function of the other $n+2$ fields,
we may rewrite the above as 
\begin{eqnarray}
 &  & \left(\frac{\partial^{2}f_{1}}{\partial\sigma^{2}}+\frac{\partial^{2}f_{1}}{\partial\sigma\partial U_{n}}\frac{\partial U_{n}}{\partial\sigma}\right)\! d\sigma+\left(\frac{\partial^{2}f_{1}}{\partial\sigma\partial U_{j}}+\frac{\partial^{2}f_{1}}{\partial\sigma\partial U_{n}}\frac{\partial U_{n}}{\partial U_{j}}\right)\! dU_{j}\nonumber \\
 &  & \qquad+\left(1+\frac{\partial^{2}f_{1}}{\partial\sigma\partial U_{n}}\frac{\partial U_{n}}{\partial\lambda_{1}}\right)\! d\lambda_{1}+\left(\frac{\partial^{2}f_{1}}{\partial\sigma\partial U_{n}}\frac{\partial U_{n}}{\partial\Phi}-1\right)\! d\Phi=0\,.
\end{eqnarray}
This constraint has solution for 
\begin{eqnarray}
\frac{\partial U_{n}}{\partial\sigma} & = & -\frac{\frac{\partial^{2}f}{\partial\sigma^{2}}}{\frac{\partial^{2}f}{\partial\sigma\partial U_{n}}}=-\frac{f_{,\sigma\sigma}}{f_{,\sigma U_{n}}}\,,\\
\frac{\partial U_{n}}{\partial U_{j}} & = & -\frac{\frac{\partial^{2}f}{\partial\sigma\partial U_{j}}}{\frac{\partial^{2}f}{\partial\sigma\partial U_{n}}}=-\frac{f_{,\sigma U_{j}}}{f_{,\sigma U_{n}}}\,,\\
\frac{\partial U_{n}}{\partial\lambda_{1}} & = & -\frac{1}{\frac{\partial^{2}f}{\partial\sigma\partial U_{n}}}=-\frac{1}{f_{,\sigma U_{n}}}\,,\\
\frac{\partial U_{n}}{\partial\Phi} & = & \frac{1}{\frac{\partial^{2}f}{\partial\sigma\partial U_{n}}}=\frac{1}{f_{,\sigma U_{n}}}\,,
\end{eqnarray}
where we have replaced $f_{1}$ by $f$ for notational simplicity,
which is allowed because $\partial f_{1}/\partial\sigma=\partial f/\partial\sigma$
by definition.

Using the above result, the action is further rewritten as 
\begin{eqnarray}
S_{m\leq n} & = & \int d^{4}x\sqrt{-g}\Biggl[\Phi R+g^{\alpha\beta}(\partial_{\alpha}\lambda_{1}\partial_{\beta}U_{1}+\partial_{\alpha}\lambda_{2}\partial_{\beta}U_{2}+\cdots+\partial_{\alpha}\lambda_{n-1}\partial_{\beta}U_{n-1})\nonumber \\
 &  & {}+g^{\alpha\beta}\partial_{\alpha}\lambda_{n}\left(\frac{1}{f_{,\sigma U_{n}}}\,\partial_{\beta}\Phi-\frac{f_{,\sigma\sigma}}{f_{,\sigma U_{n}}}\,\partial_{\beta}\sigma-\frac{f_{,\sigma U_{j}}}{f_{,\sigma U_{n}}}\,\partial_{\beta}U_{j}-\frac{1}{f_{,\sigma U_{n}}}\,\partial_{\beta}\lambda_{1}\right)\nonumber \\
 &  & {}+f_{1}(U_{1},\cdots,U_{n})-\frac{\partial f_{1}}{\partial\sigma}\sigma+\lambda_{2}U_{1}+\cdots+\lambda_{n}U_{n-1}+f_{2}(U_{1},\cdots,U_{m})\Biggr],
\end{eqnarray}
or 
\begin{eqnarray}
S_{m>n} & = & \int d^{4}x\sqrt{-g}\Biggl[\Phi R+g^{\alpha\beta}(\partial_{\alpha}\lambda_{1}\partial_{\beta}U_{1}+\partial_{\alpha}\lambda_{2}\partial_{\beta}U_{2}+\partial_{\alpha}\lambda_{n-1}\partial_{\beta}U_{n-1}+\partial_{\alpha}\lambda_{n+1}\partial_{\beta}U_{n+1}+\cdots+\partial_{\alpha}\lambda_{m}\partial_{\beta}U_{m})\nonumber \\
 &  & {}+g^{\alpha\beta}\partial_{\alpha}\lambda_{n}\left(\frac{1}{f_{,\sigma U_{n}}}\,\partial_{\beta}\Phi-\frac{f_{,\sigma\sigma}}{f_{,\sigma U_{n}}}\,\partial_{\beta}\sigma-\frac{f_{,\sigma U_{j}}}{f_{,\sigma U_{n}}}\,\partial_{\beta}U_{j}-\frac{1}{f_{,\sigma U_{n}}}\,\partial_{\beta}\lambda_{1}\right)\nonumber \\
 &  & {}+f_{1}(U_{1},\cdots,U_{n})-\frac{\partial f_{1}}{\partial\sigma}\sigma+\lambda_{2}U_{1}+\cdots+\lambda_{m}U_{m-1}+f_{2}(U_{1},\cdots,U_{m})\Biggr].
\end{eqnarray}

On performing the following conformal transformation 
\begin{eqnarray}
\bar{g}_{\alpha\beta} & = & \xi g_{\alpha\beta}\,,\\
\sqrt{-g} & = & \frac{\sqrt{-\bar{g}}}{\xi^{2}}\,,\\
R & = & \xi\left[\bar{R}+3\bar{\Box}\ln\xi-\frac{3}{2\xi^{2}}\,\bar{g}^{\alpha\beta}\partial_{\alpha}\xi\partial_{\beta}\xi\right]\,,\label{eq:conf-1-1}
\end{eqnarray}
where 
\begin{equation}
\xi\equiv\frac{2\Phi}{\Mpl^{2}}\,,
\end{equation}
we can see that the new action, for $m\leq n$, becomes, up to a total
derivative, 
\begin{eqnarray}
S_{m\leq n} & = & \int d^{4}x\sqrt{-\bar{g}}\left[\frac{\Mpl^{2}}{2}\,\bar{R}-\frac{3\Mpl^{2}}{4\Phi^{2}}\bar{g}^{\alpha\beta}\partial_{\alpha}\Phi\partial_{\beta}\Phi+\frac{\Mpl^{2}\bar{g}^{\alpha\beta}}{2\Phi}\partial_{\alpha}\lambda_{j}\partial_{\beta}U_{j}-V\right.\\
 &  & {}+\left.\frac{\Mpl^{2}\bar{g}^{\alpha\beta}}{2\Phi}\partial_{\alpha}\lambda_{n}\left(\frac{1}{f_{,\sigma U_{n}}}\partial_{\beta}\Phi-\frac{f_{,\sigma\sigma}}{f_{,\sigma U_{n}}}\partial_{\beta}\sigma-\frac{f_{,\sigma U_{j}}}{f_{,\sigma U_{n}}}\partial_{\beta}U_{j}-\frac{1}{f_{,\sigma U_{n}}}\partial_{\beta}\lambda_{1}\right)\right],
\end{eqnarray}
where $j=1,\cdots,n-1$, and 
\begin{equation}
V=-\frac{\Mpl^{4}}{4\Phi^{2}}\left[f_{1}-\frac{\partial f_{1}}{\partial\sigma}\,\sigma+\lambda_{2}U_{1}+\cdots+\lambda_{n}U_{n-1}+f_{2}\right].
\end{equation}
On the other hand, for $m>n$, we have 
\begin{eqnarray}
S_{m>n} & = & \int d^{4}x\sqrt{-\bar{g}}\left[\frac{\Mpl^{2}}{2}\,\bar{R}-\frac{3\Mpl^{2}}{4\Phi^{2}}\bar{g}^{\alpha\beta}\partial_{\alpha}\Phi\partial_{\beta}\Phi+\frac{\Mpl^{2}\bar{g}^{\alpha\beta}}{2\Phi}\partial_{\alpha}\lambda_{j}\partial_{\beta}U_{j}+\frac{\Mpl^{2}\bar{g}^{\alpha\beta}}{2\Phi}\partial_{\alpha}\lambda_{k}\partial_{\beta}U_{k}\right.\nonumber \\
 &  & {}+\left.\frac{\Mpl^{2}\bar{g}^{\alpha\beta}}{2\Phi}\partial_{\alpha}\lambda_{n}\left(\frac{1}{f_{,\sigma U_{n}}}\partial_{\beta}\Phi-\frac{f_{,\sigma\sigma}}{f_{,\sigma U_{n}}}\partial_{\beta}\sigma-\frac{f_{,\sigma U_{j}}}{f_{,\sigma U_{n}}}\partial_{\beta}U_{j}-\frac{1}{f_{,\sigma U_{n}}}\partial_{\beta}\lambda_{1}\right)-V\right],
\end{eqnarray}
where $j=1,\cdots,n-1$, $k=n+1,\cdots,m$ and 
\begin{equation}
V=-\frac{\Mpl^{4}}{4\Phi^{2}}\left[f_{1}-\frac{\partial f_{1}}{\partial\sigma}\,\sigma+\lambda_{2}U_{1}+\cdots+\lambda_{m}U_{m-1}+f_{2}\right].
\end{equation}
In the following, we will consider the two cases, $m\leq n$ and $m>n$,
separately.

\subsection{Case $\bm{m\leq n}$}

Let us make a further field redefinition by constant rescaling as
\begin{eqnarray}
\Phi & = & \Mpl\, q_{1}\,,\\
\sigma & = & \Mpl\, q_{2}\,,\\
U_{j} & = & \frac{q_{j+2}}{\Mpl^{2j-1}}\quad(j=1,\cdots,n-1)\,,\quad U_{n}=\frac{u_{n}}{\Mpl^{2n-1}}\,,\\
\lambda_{i} & = & \Mpl^{2i-1}q_{n+1+i}\quad(i=1,\cdots,n)\,,\\
f & = & \Mpl^{2}\bar{f}\,,\qquad f_{,\sigma U_{n}}=\Mpl^{2n}\bar{f}_{,q_{2}u_{n}}\,,\\
f_{,\sigma\sigma} & = & \bar{f}_{,q_{2}q_{2}}\,,\qquad f_{,\sigma U_{j}}=\Mpl^{2j}\bar{f}_{,q_{2}q_{j+2}}\,,
\end{eqnarray}
where we have included the rescaling of $U_{n}$ although it is not
an independent field. Notice also that this rescaling is not necessary
for the function $f_{2}$ as this quantity only enters in the definition
of the potential, i.e.\ it does not affect the kinetic term of any
of the fields. Thus in total we have $2n+1$ fields that we have named
$q_{l}$ where $l=1,\cdots,2n+1$. Then the action takes the following
form: 
\begin{eqnarray}
S & = & \int d^{4}x\sqrt{-\bar{g}}\left[\frac{\Mpl^{2}}{2}\,\bar{R}-\frac{1}{2}\bar{g}^{\alpha\beta}G^{kl}\partial_{\alpha}q_{k}\partial_{\beta}q_{l}-V\right],
\end{eqnarray}
where $(k,l)=1,\cdots,2n+1$, and the kinetic-term metric $G^{kl}$
is a field-dependent symmetric matrix $G^{kl}$ whose only non-zero
elements are 
\begin{eqnarray}
G^{11} & = & \frac{3\Mpl^{2}}{2q_{1}^{2}}\,,\qquad G^{1,2n+1}=-\frac{\Mpl}{2q_{1}\bar{f}_{,q_{2}u_{n}}}\,,\qquad G^{2,2n+1}=\frac{\Mpl\bar{f}_{,q_{2}q_{2}}}{2q_{1}\bar{f}_{,q_{2}u_{n}}}\,,\\
G^{j+2,j+n+1} & = & -\frac{\Mpl}{2q_{1}}\,,\qquad G^{j+2,2n+1}=\frac{\Mpl\bar{f}_{,q_{2}q_{j+2}}}{2q_{1}\bar{f}_{,q_{2}u_{n}}}\,,\qquad G^{n+2,2n+1}=\frac{\Mpl}{2q_{1}\bar{f}_{,q_{2}u_{n}}}\,.
\end{eqnarray}

Let us analyze the kinetic matrix $\bm{G}$. In order to examine whether
the fields $q$ have positive kinetic terms, we need to study whether
$\bm{G}$ is positive-definite or not. For this purpose, we notice
that it enters in the Lagrangian in the form of $\mathcal{L}\ni\bm{v}^{T}\cdot\mathbf{G}\cdot\bm{v}$
where $\bm{v}$ is a ($2n+1$)-dimensional vector. Hence it suffices
to look for a linear transformation of $\bm{v}$ in the form $\bm{v}=\mathbf{A}\cdot\bm{w}$
which diagonalizes the matrix $\mathbf{G}$. Such a transformation
is found as 
\begin{eqnarray}
v_{1} & = & w_{1}+\frac{q_{1}}{3\Mpl\bar{f}_{,q_{2}u_{n}}}\, w_{2n}-\frac{q_{1}\bar{f}_{,q_{2}q_{2}}}{6\Mpl\bar{f}_{,q_{2}q_{3}}-q_{1}}\, w_{2n+1}\,,\\
v_{2} & = & w_{2n+1}\,,\\
v_{j+2} & = & w_{2j}-\frac{1}{2}\, w_{2j+1}+\frac{\delta_{j,1}}{\bar{f}_{,q_{2}u_{n}}}\, w_{2n}-\frac{3\Mpl\bar{f}_{,q_{2}q_{2}}\delta_{j,1}}{6\Mpl\bar{f}_{,q_{2}q_{3}}-q_{1}}\, w_{2n+1}\,,\\
v_{j+n+1} & = & w_{2j}+\frac{1}{2}\, w_{2j+1}+\frac{\bar{f}_{,q_{2}q_{j+2}}}{\bar{f}_{,q_{2}u_{n}}}\, w_{2n}-\frac{3\Mpl\bar{f}_{,q_{2}q_{2}}\bar{f}_{,q_{2}q_{j+2}}}{6\Mpl\bar{f}_{,q_{2}q_{3}}-q_{1}}\, w_{2n+1}\,,\\
v_{2n+1} & = & w_{2n}-\frac{3\Mpl\bar{f}_{,q_{2}q_{2}}\bar{f}_{,q_{2}u_{n}}}{6\Mpl\bar{f}_{,q_{2}q_{3}}-q_{1}}\, w_{2n+1}\,.
\end{eqnarray}
Then the new kinetic matrix $\tilde{\mathbf{G}}=\mathbf{A}^{T}\cdot\mathbf{G}\cdot\mathbf{A}$
becomes diagonal with the elements given by 
\begin{eqnarray}
\tilde{G}^{11} & = & \frac{3\Mpl^{2}}{2q_{1}^{2}}\,,\qquad\tilde{G}^{2j,2j}=-\frac{\Mpl}{q_{1}}\,,\qquad\tilde{G}^{2j+1,2j+1}=\frac{\Mpl}{4q_{1}}\,,\\
\tilde{G}^{2n,2n} & = & \frac{6\Mpl\bar{f}_{,q_{2}q_{3}}-q_{1}}{6q_{1}\bar{f}_{,q_{2}u_{n}}^{2}}\,,\qquad\tilde{G}^{2n+1,2n+1}=-\frac{3\Mpl^{2}\bar{f}_{,q_{2}q_{2}}^{2}}{2q_{1}(6\Mpl\bar{f}_{,q_{2}q_{3}}-q_{1})}\,.
\end{eqnarray}
As clear from the above, for this theory, we conclude that there always
exist $n$ ghosts independently of the sign of $q_{1}$.

\subsection{Case $m>n$}

In this case we perform the field redefinition, 
\begin{eqnarray}
\Phi & = & \Mpl\, q_{1}\,,\\
\sigma & = & \Mpl\, q_{2}\,,\\
U_{j} & = & \frac{q_{j+2}}{\Mpl^{2j-1}}\quad(j=1,\cdots,n-1)\,,U_{n}=\frac{u_{n}}{\Mpl^{2n-1}}\,,\\
U_{r} & = & \frac{q_{r+1}}{\Mpl^{2r-1}}\quad(r=n+1,\cdots,m)\,,\\
\lambda_{i} & = & \Mpl^{2i-1}q_{m+1+i}\quad(i=1,\cdots,m)\,,\\
f & = & \Mpl^{2}\bar{f}\,,\qquad f_{,\sigma U_{n}}\equiv\Mpl^{2n}\bar{f}_{,q_{2}u_{n}}\,,\\
f_{,\sigma\sigma} & = & \bar{f}_{,q_{2}q_{2}}\,,\qquad f_{,\sigma U_{j}}=\Mpl^{2j}\bar{f}_{,q_{2}q_{j+2}}\,,
\end{eqnarray}
where again we have rescaled the dependent field $U_{n}$ as well.
Thus we have $2m+1$ fields in total, and the action takes the form,
\begin{eqnarray}
S & = & \int d^{4}x\sqrt{-\bar{g}}\left[\frac{\Mpl^{2}}{2}\,\bar{R}-\frac{1}{2}\bar{g}^{\alpha\beta}G^{kl}\partial_{\alpha}q_{k}\partial_{\beta}q_{l}-V\right],
\end{eqnarray}
where $(k.l)=1,\cdots,2m+1$. The only non-zero elements of the kinetic-term
metric $G^{kl}$ are 
\begin{eqnarray}
G^{11} & = & \frac{3\Mpl^{2}}{2q_{1}^{2}}\,,\qquad G^{1,m+n+1}=-\frac{\Mpl}{2q_{1}\bar{f}_{,q_{2}u_{n}}}\,,\qquad G^{2,m+n+1}=\frac{\Mpl\bar{f}_{,q_{2}q_{2}}}{2q_{1}\bar{f}_{,q_{2}u_{n}}}\,,\\
G^{j+2,m+j+1} & = & -\frac{\Mpl}{2q_{1}}\,,\qquad G^{j+2,m+n+1}=\frac{\Mpl\bar{f}_{,q_{2}q_{j+2}}}{2q_{1}\bar{f}_{,q_{2}u_{n}}}\,,\qquad G^{m+2,m+n+1}=\frac{\Mpl}{2q_{1}\bar{f}_{,q_{2}u_{n}}}\,,\\
G^{r+1,m+r+1} & = & -\frac{\Mpl}{2q_{1}}\,,
\end{eqnarray}
where $j=1,\cdots,n-1$ and $r=n+1,\cdots,m$.

Once again the kinetic matrix enters the Lagrangian in the form $\bm{v}^{T}\cdot\mathbf{G}\cdot\bm{v}$,
and whether it is positive definite or not can be examined by diagonalizing
the matrix by a transformation of the form $\bm{v}=\mathbf{A}\cdot\bm{w}$.
We find that the transformation, 
\begin{eqnarray}
v_{1} & = & w_{1}+\frac{q_{1}}{3\Mpl\bar{f}_{,q_{2}u_{n}}}\, w_{2m}-\frac{q_{1}\bar{f}_{,q_{2}q_{2}}}{6\Mpl\bar{f}_{,q_{2}q_{3}}-q_{1}}\, w_{2m+1}\,,\\
v_{2} & = & w_{2m+1}\,,\\
v_{j+2} & = & w_{2j}-\frac{1}{2}\, w_{2j+1}+\frac{\delta_{j,1}}{\bar{f}_{,q_{2}u_{n}}}\, w_{2m}-\frac{3\Mpl\bar{f}_{,q_{2}q_{2}}\delta_{j,1}}{6\Mpl\bar{f}_{,q_{2}q_{3}}-q_{1}}\, w_{2m+1}\,,\\
v_{k+1} & = & w_{2k-2}-\frac{1}{2}\, w_{2k-1}\,,\\
v_{m+j+1} & = & w_{2j}+\frac{1}{2}\, w_{2j+1}+\frac{\bar{f}_{,q_{2}q_{j+2}}}{\bar{f}_{,q_{2}u_{n}}}\, w_{2m}-\frac{3\Mpl\bar{f}_{,q_{2}q_{2}}\bar{f}_{,q_{2}q_{j+2}}}{6\Mpl\bar{f}_{,q_{2}q_{3}}-q_{1}}\, w_{2m+1}\,,\\
v_{m+n+1} & = & w_{2m}-\frac{3\Mpl\bar{f}_{,q_{2}q_{2}}\bar{f}_{,q_{2}u_{n}}}{6\Mpl\bar{f}_{,q_{2}q_{3}}-q_{1}}\, w_{2m+1}\,,\\
v_{m+k+1} & = & w_{2k-2}+\frac{1}{2}\, w_{2k-1}\,,
\end{eqnarray}
diagonalizes the matrix $\bm{G}$. The new kinetic matrix $\tilde{\mathbf{G}}=\mathbf{A}^{T}\cdot\mathbf{G}\cdot\mathbf{A}$
becomes diagonal with the elements given by 
\begin{eqnarray}
\tilde{G}^{11} & = & \frac{3\Mpl^{2}}{2q_{1}^{2}}\,,\qquad\tilde{G}^{2j,2j}=-\frac{\Mpl}{q_{1}}\,,\qquad\tilde{G}^{2j+1,2j+1}=\frac{\Mpl}{4q_{1}}\,,\\
\tilde{G}^{2k-2,2k-2} & = & -\frac{\Mpl}{q_{1}}\,,\qquad\tilde{G}^{2k-1,2k-1}=\frac{\Mpl}{4q_{1}}\,,\\
\tilde{G}^{2m,2m} & = & \frac{6\Mpl\bar{f}_{,q_{2}q_{3}}-q_{1}}{6q_{1}\bar{f}_{,q_{2}u_{n}}^{2}}\,,\qquad\tilde{G}^{2m+1,2m+1}=-\frac{3\Mpl^{2}\bar{f}_{,q_{2}q_{2}}^{2}}{2q_{1}(6\Mpl\bar{f}_{,q_{2}q_{3}}-q_{1})}\,.
\end{eqnarray}
Therefore, similar to the previous case, there always exist $m$ ghosts
independently of the sign of $q_{1}$. \vspace{5mm}

To summarize, for the Lagrangian of the form (\ref{eq:lagrOR-2}),
there always exist $\mathrm{max}(n,m)$ ghosts provided $f_{1}$ is
nonlineaer in the Ricci scalar $R$.

\subsection{Case $n=0$}

In this case we want to discuss here the model described by the Lagrangian
(\ref{eq:lagrNEW-2b}) where $n=0$, that is the action can be written
as 
\begin{eqnarray}
S_{n=0} & = & \int d^{4}x\sqrt{-g}\Biggl[f_{1}(\sigma)+\frac{\partial f_{1}}{\partial\sigma}\,(R-\sigma)+f_{2}(U_{1},\cdots,U_{m})\nonumber \\
 &  & \qquad+\lambda_{1}(R-\Box U_{1})+\lambda_{2}(U_{1}-\Box U_{2})+\cdots+\lambda_{m}(U_{m-1}-\Box U_{m})\Biggr],\label{eq:lagrNEW-2b-1}
\end{eqnarray}
and we will assume 
\begin{equation}
\frac{\partial f_{2}}{\partial U_{m}}\neq0\,,\qquad\textrm{and}\qquad\frac{\partial^{2}f_{1}}{\partial\sigma^{2}}\neq0.\label{eq:condiz1}
\end{equation}
which can be described as a non-local term correction to an $f_{2}(R)$
gravity theory. The term $f_{2}$ alone is known not to introduce
any ghosts, provided that $\partial f_{1}/\partial R>0$, i.e.\ $\partial f_{1}/\partial\sigma>0$.
We can rewrite the action in the following form 
\begin{equation}
S_{n=0}=\int d^{4}x\sqrt{-g}\left[\left(\lambda_{1}+\frac{\partial f_{2}}{\partial\sigma}\right)\, R+f_{1}+f_{2}-\frac{\partial f_{2}}{\partial\sigma}\,\sigma-\lambda_{1}\Box U_{1}+\lambda_{2}(U_{1}-\Box U_{2})+\cdots+\lambda_{m}(U_{m-1}-\Box U_{m})\right],
\end{equation}
and define 
\begin{equation}
\Phi=\lambda_{1}+\frac{\partial f_{2}}{\partial\sigma}\,,
\end{equation}
which, because of Eq.\ (\ref{eq:condiz1}), can be inverted for the
field $\sigma$, as $\sigma=\sigma(\Phi,\lambda_{1})$, so that the
action becomes 
\begin{eqnarray}
S_{n=0} & = & \int d^{4}x\sqrt{-g}\,[\Phi\, R+\nabla_{\alpha}\lambda_{1}\nabla^{\alpha}U_{1}+\nabla_{\alpha}\lambda_{2}\nabla^{\alpha}U_{2}+\cdots+\nabla_{\alpha}\lambda_{m}\nabla^{\alpha}U_{m}\nonumber \\
 &  & {}+f_{1}+f_{2}\bigl(\sigma(\Phi,\lambda_{1})\bigr)-(\Phi-\lambda_{1})\,\sigma(\Phi,\lambda_{1})+\lambda_{2}U_{1}+\dots+\lambda_{m}U_{m-1}]\,.
\end{eqnarray}
On performing the conformal transformation we find 
\begin{equation}
S_{n=0}=\int d^{4}x\sqrt{-\bar{g}}\left[\frac{\Mpl^{2}}{2}\,\bar{R}-\frac{3\Mpl^{2}}{4\Phi^{2}}\bar{g}^{\alpha\beta}\partial_{\alpha}\Phi\partial_{\beta}\Phi+\frac{\Mpl^{2}\bar{g}^{\alpha\beta}}{2\Phi}\sum_{j=1}^{m}\partial_{\alpha}\lambda_{j}\partial_{\beta}U_{j}-V\right],
\end{equation}
where the potential $V$ is defined as 
\begin{equation}
V=\frac{\Mpl^{4}}{4\Phi^{2}}\left[(\Phi-\lambda_{1})\,\sigma(\Phi,\lambda_{1})-f_{1}-f_{2}-\lambda_{2}U_{1}-\dots-\lambda_{m}U_{m-1}\right].
\end{equation}

Also in this case, we can perform the following field redefinition,
\begin{eqnarray}
\Phi & = & \Mpl\, q_{1}\,,\\
U_{j} & = & \frac{q_{j+1}}{\Mpl^{2j-1}}\quad(j=1,\cdots,m)\,,\\
\lambda_{j} & = & \Mpl^{2j-1}q_{m+1+j}\quad(j=1,\cdots,m)\,,
\end{eqnarray}
so that the action becomes 
\begin{equation}
S_{n=0}=\int d^{4}x\sqrt{-\bar{g}}\left[\frac{\Mpl^{2}}{2}\,\bar{R}-\frac{3\Mpl^{2}}{4q_{1}^{2}}\bar{g}^{\alpha\beta}\partial_{\alpha}q_{1}\partial_{\beta}q_{1}+\frac{\Mpl\bar{g}^{\alpha\beta}}{2q_{1}}\sum_{j=1}^{m}\partial_{\alpha}q_{m+1+j}\partial_{\beta}q_{j+1}-V\right],
\end{equation}
or, the equivalent form, 
\begin{eqnarray}
S & = & \int d^{4}x\sqrt{-\bar{g}}\left[\frac{\Mpl^{2}}{2}\,\bar{R}-\frac{1}{2}\bar{g}^{\alpha\beta}G^{kl}\partial_{\alpha}q_{k}\partial_{\beta}q_{l}-V\right],
\end{eqnarray}
where $(k.l)=1,\cdots,2m+1$. The only non-zero elements of the symmetric
kinetic-term metric $G^{kl}$ are 
\begin{eqnarray}
G^{11} & = & \frac{3\Mpl^{2}}{2q_{1}^{2}}\,,\qquad G^{j+1,j+1+m}=-\frac{\Mpl}{2q_{1}}\,,
\end{eqnarray}
where $j=1,\cdots,m$. On making the following final field redefinition
\begin{equation}
q_{1}=Q_{1}\,,\qquad q_{j+1}=Q_{2j}-\frac{Q_{2j+1}}{2}\,,\qquad q_{j+1+m}=Q_{2j}+\frac{Q_{2j+1}}{2}\,,
\end{equation}
then the new kinetic matrix becomes diagonal with elements 
\begin{equation}
\hat{G}^{11}=\frac{3\Mpl^{2}}{2q_{1}^{2}}\,,\qquad\hat{G}^{2j,2j}=-\frac{\Mpl}{q_{1}}\,,\qquad\hat{G}^{2j+1,2j+1}=\frac{\Mpl}{4q_{1}}\,.
\end{equation}
Therefore, for this theory, independently of the sign of $q_{1}$,
there will always exist $m$ ghosts, independently of the sign of
$q_{1}$.

\section{Linear-$R$ non-local gravity theory}

\label{sec:linear}

Let us consider a subcase of the theory where the Lagrangian is linear
in $R$, which corresponds to the case $\partial^{2}f_{1}/\partial\sigma^{2}=0$
in the action (\ref{eq:lagrOR-2}) studied in the previous section.
Namely we consider 
\begin{equation}
S=\int d^{4}x\sqrt{-g}\,[R\, f_{1}(\Box^{-1}R,\Box^{-2}R,\cdots,\Box^{-n}R)+f_{2}(\Box^{-1}R,\Box^{-2}R,\cdots,\Box^{-m}R)]\,,\label{eq:lagrOR-1}
\end{equation}
where we suppose $n\geq2$. For completeness, the special case $n=1$,
which has been already studied in the literature, will be discussed
separately.

We rewrite the action, along the same lines as the previous section,
as 
\begin{eqnarray}
S_{m\leq n} & = & \int d^{4}x\sqrt{-g}\,[Rf_{1}(U_{1},\cdots,U_{n})+\lambda_{1}(R-\Box U_{1})+\lambda_{2}(U_{1}-\Box U_{2})+\cdots\nonumber \\
 &  & +\lambda_{n}(U_{n-1}-\Box U_{n})+f_{2}(U_{1},\cdots,U_{m})]\,,\label{eq:lagrNEW-1}
\end{eqnarray}
for $m\leq n$, and 
\begin{eqnarray}
S_{m>n} & = & \int d^{4}x\sqrt{-g}\,[Rf_{1}(U_{1},\cdots,U_{n})+\lambda_{1}(R-\Box U_{1})+\lambda_{2}(U_{1}-\Box U_{2})+\cdots\nonumber \\
 &  & +\lambda_{n}(U_{n-1}-\Box U_{n})+\cdots+\lambda_{m}(U_{m-1}-\Box U_{m})+f_{2}(U_{1},\cdots,U_{m})]\,,
\end{eqnarray}
for $m>n$. These can be cast into the form, 
\begin{eqnarray}
S_{m\leq n} & = & \int d^{4}x\sqrt{-g}\left[(f_{1}+\lambda_{1})\, R+g^{\alpha\beta}(\partial_{\alpha}\lambda_{1}\partial_{\beta}U_{1}+\partial_{\alpha}\lambda_{2}\partial_{\beta}U_{2}+\cdots+\partial_{\alpha}\lambda_{n}\partial_{\beta}U_{n})\right.\nonumber \\
 &  & +\left.\lambda_{2}U_{1}+\cdots+\lambda_{n}U_{n-1}+f_{2}(U_{1},\cdots,U_{m})\right],
\end{eqnarray}
and 
\begin{eqnarray}
S_{m>n} & = & \int d^{4}x\sqrt{-g}\left[(f_{1}+\lambda_{1})\, R+g^{\alpha\beta}(\partial_{\alpha}\lambda_{1}\partial_{\beta}U_{1}+\cdots+\partial_{\alpha}\lambda_{n}\partial_{\beta}U_{n}+\cdots+\partial_{\alpha}\lambda_{m}\partial_{\beta}U_{m})\right.\nonumber \\
 &  & +\left.\lambda_{2}U_{1}+\cdots+\lambda_{m}U_{m-1}+f_{2}(U_{1},\cdots,U_{m})\right],
\end{eqnarray}
respectively.

Let us make the field redefinition as 
\begin{equation}
f_{1}(U_{1},\cdots,U_{n})+\lambda_{1}=\Phi\,,\label{eq:constrU}
\end{equation}
and let us use this equation to express $U_{n}$ in terms of the other
fields as 
\begin{equation}
U_{n}=U_{n}(\Phi-\lambda_{1},U_{j})\,;\qquad j=1,\cdots,n-1\,.
\end{equation}
In this case the derivative of Eq.~(\ref{eq:constrU}) gives 
\begin{equation}
\frac{\partial f_{1}}{\partial U_{j}}\, dU_{j}+\frac{\partial f_{1}}{\partial U_{n}}\, dU_{n}+d\lambda_{1}-d\Phi=0\,,
\end{equation}
or 
\begin{equation}
\left(\frac{\partial f_{1}}{\partial U_{j}}+\frac{\partial f_{1}}{\partial U_{n}}\frac{\partial U_{n}}{\partial U_{j}}\right)\, dU_{j}+\left(1+\frac{\partial f_{1}}{\partial U_{n}}\frac{\partial U_{n}}{\partial\lambda_{1}}\right)d\lambda_{1}+\left(\frac{\partial f_{1}}{\partial U_{n}}\frac{\partial U_{n}}{\partial\Phi}-1\right)d\Phi=0\,.
\end{equation}
This implies 
\begin{eqnarray}
\frac{\partial U_{n}}{\partial U_{j}} & = & -\frac{\frac{\partial f_{1}}{\partial U_{j}}}{\frac{\partial f_{1}}{\partial U_{n}}}=-\frac{f_{1,U_{j}}}{f_{1,U_{n}}}\,,\\
\frac{\partial U_{n}}{\partial\lambda_{1}} & = & -\frac{1}{f_{1,U_{n}}}\,,\\
\frac{\partial U_{n}}{\partial\Phi} & = & \frac{1}{f_{1,U_{n}}}\,.
\end{eqnarray}
Therefore, the actions for $m\leq n$ and $m>n$ can be rewritten,
respectively, as 
\begin{eqnarray}
S_{m\leq n} & = & \int d^{4}x\sqrt{-g}\Biggl[\Phi R+g^{\alpha\beta}\Biggl(\partial_{\alpha}\lambda_{1}\partial_{\beta}U_{1}+\cdots+\partial_{\alpha}\lambda_{n-1}\partial_{\beta}U_{n-1}-\frac{f_{1,j}}{f_{1,n}}\,\partial_{\alpha}\lambda_{n}\partial_{\beta}U_{j}\nonumber \\
 &  & -\frac{1}{f_{1,n}}\,\partial_{\alpha}\lambda_{n}\partial_{\beta}\lambda_{1}+\frac{1}{f_{1,n}}\,\partial_{\alpha}\lambda_{n}\partial_{\beta}\Phi\Biggr)+\lambda_{2}U_{1}+\cdots+\lambda_{n}U_{n-1}\Biggr],
\end{eqnarray}
and 
\begin{eqnarray}
S_{m>n} & = & \int d^{4}x\sqrt{-g}\Biggl[\Phi R+g^{\alpha\beta}\Biggl(\partial_{\alpha}\lambda_{1}\partial_{\beta}U_{1}+\cdots+\partial_{\alpha}\lambda_{n-1}\partial_{\beta}U_{n-1}-\frac{f_{1,j}}{f_{1,n}}\,\partial_{\alpha}\lambda_{n}\partial_{\beta}U_{j}\nonumber \\
 &  & +\partial_{\alpha}\lambda_{n+1}\partial_{\beta}U_{n+1}+\cdots+\partial_{\alpha}\lambda_{m}\partial_{\beta}U_{m}-\frac{1}{f_{1,n}}\,\partial_{\alpha}\lambda_{n}\partial_{\beta}\lambda_{1}+\frac{1}{f_{1,n}}\,\partial_{\alpha}\lambda_{n}\partial_{\beta}\Phi\Biggr)\nonumber \\
 &  & +\lambda_{2}U_{1}+\cdots+\lambda_{n+1}U_{n}(U_{j},\Phi-\lambda_{1})+\cdots+\lambda_{m}U_{m-1}+f_{2}(U_{1},\cdots,U_{m})\Biggr].
\end{eqnarray}

Let us now perform a conformal transformation to the Einstein frame,
as in the previous section. For $m\leq n$ we find 
\begin{eqnarray}
S_{m\leq n} & = & \int d^{4}x\sqrt{-\bar{g}}\Biggl[\frac{\Mpl^{2}}{2}\,\bar{R}-\frac{3\Mpl^{2}}{4\Phi^{2}}\bar{g}^{\alpha\beta}\partial_{\alpha}\Phi\partial_{\beta}\Phi+\frac{\Mpl^{2}\bar{g}^{\alpha\beta}}{2\Phi}\Biggl(\partial_{\alpha}\lambda_{1}\partial_{\beta}U_{1}+\partial_{\alpha}\lambda_{2}\partial_{\beta}U_{2}+\cdots+\partial_{\alpha}\lambda_{n-1}\partial_{\beta}U_{n-1}\nonumber \\
 &  & -\frac{f_{1,U_{j}}}{f_{1,U_{n}}}\,\partial_{\alpha}\lambda_{n}\partial_{\beta}U_{j}-\frac{1}{f_{1,U_{n}}}\,\partial_{\alpha}\lambda_{n}\partial_{\beta}\lambda_{1}+\frac{1}{f_{1,U_{n}}}\,\partial_{\alpha}\lambda_{n}\partial_{\beta}\Phi\Biggr)-V_{1}\Biggr],
\end{eqnarray}
where 
\begin{equation}
V_{1}=-\frac{\Mpl^{4}}{4\Phi^{2}}\,[\lambda_{2}U_{1}+\cdots+\lambda_{n}U_{n-1}+f_{2}]\,.
\end{equation}
For $m>n$ we find 
\begin{eqnarray}
S_{m>n} & = & \int d^{4}x\sqrt{-\bar{g}}\Biggl[\frac{\Mpl^{2}}{2}\,\bar{R}-\frac{3\Mpl^{2}}{4\Phi^{2}}\bar{g}^{\alpha\beta}\partial_{\alpha}\Phi\partial_{\beta}\Phi+\frac{\Mpl^{2}\bar{g}^{\alpha\beta}}{2\Phi}\Biggl(\partial_{\alpha}\lambda_{1}\partial_{\beta}U_{1}+\partial_{\alpha}\lambda_{2}\partial_{\beta}U_{2}\nonumber \\
 &  & +\cdots+\partial_{\alpha}\lambda_{n-1}\partial_{\beta}U_{n-1}-\frac{f_{1,U_{j}}}{f_{1,U_{n}}}\,\partial_{\alpha}\lambda_{n}\partial_{\beta}U_{j}-\frac{1}{f_{1,U_{n}}}\,\partial_{\alpha}\lambda_{n}\partial_{\beta}\lambda_{1}\nonumber \\
 &  & +\frac{1}{f_{1,U_{n}}}\,\partial_{\alpha}\lambda_{n}\partial_{\beta}\Phi+\partial_{\alpha}\lambda_{n+1}\partial_{\beta}U_{n+1}+\cdots+\partial_{\alpha}\lambda_{m}\partial_{\beta}U_{m}\Biggr)-V_{2}\Biggr],
\end{eqnarray}
where 
\begin{equation}
V_{2}=-\frac{\Mpl^{4}}{4\Phi^{2}}\,[\lambda_{2}U_{1}+\cdots+\lambda_{m}U_{m-1}+f_{2}]\,.
\end{equation}
We are now ready to discuss the problem. As before we consider the
two cases, $m\leq n$ and $m>n$, separately.

\subsection{Case $m\leq n$}

We perform the field redefinition, 
\begin{eqnarray}
\Phi & = & \Mpl\, q_{1}\,,\\
U_{j} & = & \frac{q_{j+1}}{\Mpl^{2j-1}}\quad(j=1,\cdots,n-1)\,,\qquad U_{n}=\frac{u_{n}}{\Mpl^{2n-1}}\,,\\
\lambda_{i} & = & \Mpl^{2i-1}q_{n+i}\quad(i=1,\cdots,n)\,,\\
f_{1} & = & \Mpl\,\bar{f}_{1}\,,\\
f_{1,U_{n}} & \equiv & \Mpl^{2n}\bar{f}_{1,u_{n}}\,,\\
f_{1,U_{j}} & = & \Mpl^{2j}\bar{f}_{1,q_{j+1}}\,.
\end{eqnarray}
Note that $u_{n}$ is a function of the other $q_{k}$ fields.
Then the Lagrangian becomes 
\begin{eqnarray}
S & = & \int d^{4}x\sqrt{-\bar{g}}\left[\frac{\Mpl^{2}}{2}\,\bar{R}-\frac{1}{2}\bar{g}^{\alpha\beta}G^{kl}\partial_{\alpha}q_{k}\partial_{\beta}q_{l}-V\right],
\end{eqnarray}
where $k,l=1,\cdots,2n$. Therefore this Lagrangian has one degree
of freedom less than the previous general case, as expected.

The non-vanishing elements of the kinetic (symmetric) matrix $G^{kl}$
are 
\begin{eqnarray}
G^{11} & = & \frac{3\Mpl^{2}}{2q_{1}^{2}}\,,\qquad G^{1,2n}=-\frac{\Mpl}{2q_{1}\bar{f}_{1,u_{n}}}\,,\qquad G^{j+1,j+n}=-\frac{\Mpl}{2q_{1}}\,,\qquad G^{j+1,2n}=\frac{\Mpl}{2q_{1}}\frac{\bar{f}_{1,q_{j+1}}}{\bar{f}_{1,u_{n}}}\,,\\
G^{n+1,2n} & = & \frac{\Mpl}{2q_{1}\bar{f}_{1,u_{n}}}\,.
\end{eqnarray}
As before we examine whether $\bm{v}^{T}\cdot\mathbf{G}\cdot\bm{v}$
for a general $2n$-dimensional vector $\bm{v}$ is positive definite
or not. We find that a transformation of the form $\bm{v}=\mathbf{A}\cdot\bm{w}$
that diagonalizes the matrix $\mathbf{G}$ is given by 
\begin{eqnarray}
v_{1} & = & w_{1}+\frac{q_{1}}{3\Mpl}\,\frac{1}{\bar{f}_{1,u_{n}}}\, w_{2n}\,,\\
v_{j+1} & = & w_{2j}-\frac{1}{2}\, w_{2j+1}+\frac{\delta_{j,1}}{\bar{f}_{1,u_{n}}}\, w_{2n}\,,\\
v_{j+n} & = & w_{2j}+\frac{1}{2}\, w_{2j+1}+\frac{\bar{f}_{1,q_{j+1}}}{\bar{f}_{1,u_{n}}}\, w_{2n}\,,\\
v_{2n} & = & w_{2n}\,.
\end{eqnarray}
In fact in this case the new kinetic matrix $\tilde{\mathbf{G}}\equiv\mathbf{A}^{T}\cdot\mathbf{G}\cdot\mathbf{A}$,
is diagonal with the non-zero elements given by 
\begin{equation}
\tilde{G}^{11}=\frac{3\Mpl^{2}}{2q_{1}^{2}}\,,\qquad\tilde{G}^{2j,2j}=-\frac{\Mpl}{q_{1}}\,,\qquad\tilde{G}^{2j+1,2j+1}=\frac{\Mpl}{4q_{1}}\,,\qquad\tilde{G}^{2n,2n}=\frac{1}{\bar{f}_{1,u_{n}}^{2}}\left[\frac{\Mpl}{q_{1}}\,\bar{f}_{1,q_{j+1}}-\frac{1}{6}\right]\,.
\end{equation}
Thus we conclude that this theory contains, in general, at least $n-1$
ghosts independently of the sign of $q_{1}$.

\subsection{Case $m>n$}

In this case, we perform the field redefinition, 
\begin{eqnarray}
\Phi & = & \Mpl\, q_{1}\,,\\
U_{j} & = & \frac{q_{j+1}}{\Mpl^{2j-1}}\quad(j=1,\cdots,n-1)\,,\qquad U_{n}=\frac{u_{n}}{\Mpl^{2n-1}}\,,\\
U_{r} & = & \frac{q_{r}}{\Mpl^{2r-1}}\quad(r=n+1,\cdots,m)\,,\\
\lambda_{i} & = & \Mpl^{2i-1}q_{m+i}\quad(i=1,\cdots,m)\,,\\
f_{1} & = & \Mpl\bar{f}_{1}\,,\qquad f_{1,U_{n}}=\Mpl^{2n}\bar{f}_{1,u_{n}}\,,\\
f_{1,U_{j}} & = & \Mpl^{2j}\bar{f}_{1,q_{j+1}}\,,
\end{eqnarray}
where $u_{n}$ is a function of the other fields. In total, there
are $2m$ fields. The kinetic matrix for the scalar fields is in the
form, 
\begin{eqnarray}
S & = & \int d^{4}x\sqrt{-\bar{g}}\left[\frac{\Mpl^{2}}{2}\,\bar{R}-\frac{1}{2}\bar{g}^{\alpha\beta}G^{kl}\partial_{\alpha}q_{k}\partial_{\beta}q_{l}-V\right],
\end{eqnarray}
where $k,l=1,\cdots,2m$. The non-zero elements of the matrix $G^{kl}$
are 
\begin{eqnarray}
G^{11} & = & \frac{3\Mpl^{2}}{2q_{1}^{2}}\,,\qquad G^{1,m+n}=-\frac{\Mpl}{2q_{1}\bar{f}_{,q_{2}u_{n}}}\,,\\
G^{j+1,m+j} & = & -\frac{\Mpl}{2q_{1}}\,,\qquad G^{j+1,m+n}=\frac{\Mpl\bar{f}_{,q_{j+1}}}{2q_{1}\bar{f}_{,u_{n}}}\,,\qquad G^{m+1,m+n}=\frac{\Mpl}{2q_{1}\bar{f}_{,q_{2}u_{n}}}\,,\\
G^{k,m+k} & = & -\frac{\Mpl}{2q_{1}}\,,
\end{eqnarray}
where $j=1,\cdots,n-1$, $k=n+1,\cdots,m$.

This time, the transformation $\bm{v}=\bm{A}\bm{w}$ that diagonalizes
the matrix $\bm{G}$ is 
\begin{eqnarray}
v_{1} & = & w_{1}+\frac{q_{1}}{3\Mpl\bar{f}_{,1u_{n}}}\, w_{2m}\,,\\
v_{j+1} & = & w_{2j}-\frac{1}{2}\, w_{2j+1}+\frac{\delta_{j,1}}{\bar{f}_{,1u_{n}}}\, w_{2m}\,,\\
v_{k} & = & w_{2k-2}-\frac{1}{2}\, w_{2k-1}\,,\\
v_{m+j} & = & w_{2j}+\frac{1}{2}\, w_{2j+1}+\frac{\bar{f}_{1,q_{j+1}}}{\bar{f}_{1,u_{n}}}\, w_{2m}\,,\\
v_{m+n} & = & w_{2m}\,,\\
v_{m+k} & = & w_{2k-2}+\frac{1}{2}\, w_{2k-1}\,.
\end{eqnarray}
The new diagonal kinetic matrix $\tilde{\mathbf{G}}=\mathbf{A}^{T}\cdot\mathbf{G}\cdot\mathbf{A}$
has the elements, 
\begin{eqnarray}
\tilde{G}^{11} & = & \frac{3\Mpl^{2}}{2q_{1}^{2}}\,,\qquad\tilde{G}^{2j,2j}=-\frac{\Mpl}{q_{1}}\,,\qquad\tilde{G}^{2j+1,2j+1}=\frac{\Mpl}{4q_{1}}\,,\\
\tilde{G}^{2k-2,2k-2} & = & -\frac{\Mpl}{q_{1}}\,,\qquad\tilde{G}^{2k-1,2k-1}=\frac{\Mpl}{4q_{1}}\,,\\
\tilde{G}^{2m,2m} & = & \frac{6\Mpl\bar{f}_{1,q_{2}}-q_{1}}{6q_{1}\bar{f}_{1,u_{n}}^{2}}\,.
\end{eqnarray}
Therefore, in this case there always exist $m-1$ ghosts, independently
of the sign of $q_{1}$, \vspace{5mm}

To summarize, in the case the Lagrangian is linear in $R$, and provided
that $n\geq2$, there always exist at least $\mathrm{max}(m,n)-1$
ghosts in the theory.

\subsection{No ghost case}

The only sub-theory which can be made free from ghosts is the case
$n=1$ and $m=0$ ($f_{2}=0$). The Lagrangian in this case reads
\begin{equation}
S=\int d^{4}x\sqrt{-g}\, R\, f_{1}(\Box^{-1}R)\,,\label{eq:lagrOR-1-1}
\end{equation}
Following the same procedure used in the previous section, we can
rewrite the action as 
\begin{equation}
S=\int d^{4}x\sqrt{-g}\Bigl[Rf_{1}(U_{1})+\lambda_{1}(R-\Box U_{1})\Bigr],\label{eq:lagrNEW-1-1}
\end{equation}
or 
\begin{eqnarray}
S & = & \int d^{4}x\sqrt{-g}\Bigl[(f_{1}+\lambda_{1})\, R+g^{\alpha\beta}\partial_{\alpha}\lambda_{1}\partial_{\beta}U_{1}\Bigr]\,.\label{eq:lagre-NEW2-1}
\end{eqnarray}
Let us make the field redefinition, 
\begin{equation}
f_{1}(U_{1})+\lambda_{1}=\Phi\,,\label{eq:constrU-1}
\end{equation}
and use this equation to express $U_{1}$ in terms of the other two
fields as 
\begin{equation}
U_{1}=U_{1}(\Phi-\lambda_{1})\,.
\end{equation}
The derivative of Eq.~(\ref{eq:constrU-1}) gives 
\begin{equation}
\frac{df_{1}}{dU_{1}}\, dU_{1}+d\lambda_{1}-d\Phi=0\,,
\end{equation}
or 
\begin{equation}
\left(1+\frac{df_{1}}{dU_{1}}\frac{\partial U_{1}}{\partial\lambda_{1}}\right)d\lambda_{1}+\left(\frac{df_{1}}{dU_{1}}\frac{\partial U_{1}}{\partial\Phi}-1\right)d\Phi=0\,.
\end{equation}
This implies 
\begin{eqnarray}
\frac{\partial U_{1}}{\partial\lambda_{1}} & = & -\frac{1}{f_{1,U_{1}}}\,,\\
\frac{\partial U_{1}}{\partial\Phi} & = & \frac{1}{f_{1,U_{1}}}\,.
\end{eqnarray}
Therefore, the action (\ref{eq:lagre-NEW2-1}) can be rewritten as
\begin{eqnarray}
S & = & \int d^{4}x\sqrt{-g}\left[\Phi R+\frac{g^{\alpha\beta}}{f_{1,U_{1}}}\left(\partial_{\alpha}\lambda_{1}\partial_{\beta}\Phi-\partial_{\alpha}\lambda_{1}\partial_{\beta}\lambda_{1}\right)\right],
\end{eqnarray}
which may be transformed to the Einstein frame as 
\begin{eqnarray}
S & = & \int d^{4}x\sqrt{-\bar{g}}\left[\frac{\Mpl^{2}}{2}\,\bar{R}-\frac{3\Mpl^{2}}{4\Phi^{2}}\bar{g}^{\alpha\beta}\partial_{\alpha}\Phi\partial_{\beta}\Phi+\frac{\Mpl^{2}\bar{g}^{\alpha\beta}}{2\Phi f_{1,U_{1}}}\left(\partial_{\alpha}\lambda_{1}\partial_{\beta}\Phi-\partial_{\alpha}\lambda_{1}\partial_{\beta}\lambda_{1}\right)\right].
\end{eqnarray}

We can still perform the field redefinition, 
\begin{eqnarray}
\Phi & = & \Mpl\, q_{1}\,,\\
\lambda_{1} & = & \Mpl q_{2}\,,\\
f_{1} & = & \Mpl\,\bar{f}_{1}\,.
\end{eqnarray}
Together with $U_{1}=u_{1}/\Mpl$, we then find 
\begin{equation}
f_{1,U_{1}}=\frac{df_{1}(U_{1})}{dU_{1}}=\Mpl^{2}\frac{d\bar{f}_{1}}{du_{1}}=\Mpl^{2}\,\bar{f}_{1,u_{1}}\,,
\end{equation}
where $u_{1}$ is a function of a linear combination $q_{1}-q_{2}$,
$u_{1}=u_{1}(q_{1}-q_{2})$\,. The action now takes the form, 
\begin{equation}
S=\int d^{4}x\sqrt{-\bar{g}}\left[\frac{\Mpl^{2}}{2}\,\bar{R}-\frac{1}{2}\, G^{ij}\bar{g}^{\alpha\beta}\partial_{\alpha}q_{i}\partial_{\beta}q_{j}\right],
\end{equation}
where $i,j=1,2$, and 
\[
\mathbf{G}=\left(\begin{array}{cc}
\frac{3\Mpl^{2}}{2q_{1}^{2}} & -\frac{\Mpl}{2q_{1}\bar{f}_{1,u_{1}}}\\
-\frac{\Mpl}{2q_{1}\bar{f}_{1,u_{1}}} & \frac{\Mpl}{q_{1}\bar{f}_{1,u_{1}}}
\end{array}\right).
\]
This matrix is positive definite when 
\begin{equation}
\frac{6\Mpl}{q_{1}}\,\bar{f}_{1,u_{1}}>1\,.
\end{equation}
Only when this condition is satisfied, the theory can be made free
from ghost \cite{Ying-li}.

\section{Conclusion}

\label{sec:conclusion}

We considered a class of non-local gravity where the Lagrangian is
a general function of $\Box^{-k}R$ ($k=1,2,\cdots,n$) where $R$
is the Ricci scalar, and studied its formally equivalent local Lagrangian
by introducing auxiliary fields. Taking the viewpoint that the physical
degrees of freedom in thus localized theory properly represent those
in the original non-local theory, we examined the kinetic term of
the localized Lagrangian to see whether there are ghosts or not.

We found that, for a theory which contains a nonlinear function of
$R$, there always exist $n$ ghost fields for $n\geq1$, while for
a theory linear in $R$, there always exist $n-1$ ghost fields for
$n\geq2$. The case of $n=1$ with linear $R$ has been already studied
and it is known that one may or may not make the theory ghost-free
depending on the choice of the parameters.

Thus except for the special case, this class of non-local gravity
always suffers from the presence of a ghost. This result makes these
theories problematic to be used as effective theories to describe
the evolution of the universe at all times. The only possible way-out
seems to be the case when the masses of these ghosts are individually
tuned to be larger than the cut-off of the theory, e.g.\ larger than
the Planck mass (in some other contexts, the cut-off mass can be lowered,
but a careful choice of the functions $f_{1}$ and/or $f_{2}$ is
needed in order to achieve large masses).

Of course, if we abandon our localization approach used to discuss
the degrees of freedom, this ghost proliferation may be avoided. For
example, one could regard the operator $\Box^{-1}$ in the Lagrangian
as $\Box_{\mathrm{ret}}^{-1}$. One would need a new formalism to
deal with such a Lagrangian. In particular, it is not clear at all
how to quantum the theory in this case. After all, a ghost field is
fatally dangerous in quantum theory. Thus avoiding ghosts by using
$\Box_{\mathrm{ret}}^{-1}$ in the Lagrangian may simply mean avoiding
quantization.

Our result leads to discussion on the fundamentals of theories of
non-local gravity with $\Box^{-k}R$ ($k=1,2,\cdots,n$), about how
it is possible to make sense of it in terms of understanding the physical
degrees of freedom. We hope this discussion will stimulate other studies
on conditions for healthy extensions of a field theory and the quantization
procedure for such non-standard Lagrangians.

\begin{acknowledgments}
We thank Ying-li Zhang for fruitful discussions. This work was supported in part by the JSPS Grant-in-Aid for
Scientific Research (A) No.\ 21244033. The work was partially performed during the workshop YITP-T-14-04.
\end{acknowledgments}

\end{document}